\def\ts     {\thinspace}
\def\kms    {\ifmmode{{\rm \ts km\ts s}^{-1}}\else{\ts km\ts s$^{-1}$}\fi}
\def\mo     {M$_{\odot}$}
\def\mue    {$\mu$m}

\documentclass{aa}
\usepackage{graphics} 

\begin{document}
  \title{Small-area molecular structures without shielding}
  \author{Andreas Heithausen}
  \institute{ 
Radioastronomisches Institut der Universit\"at Bonn, Auf dem H\"ugel 71,
53121 Bonn, Germany
}
  \offprints { A. Heithausen\\
E-mail: heith@astro.uni-bonn.de }
  \date {Received 25 June 2002, Accepted 20 August 2002}
\authorrunning{Heithausen }
\titlerunning{Molecular structures without shielding}

\abstract{
Using the IRAM 30\,m telescope two molecular structures have been  detected
which cover very small areas, $FWHM\le1'$. The clouds have velocities of
$v_{\rm lsr}\approx 5$\,\kms\ and linewidth of $\Delta v\approx0.8$\,\kms;
thus they belong most likely to the Milky Way. Applying standard conversion
factors one finds that even at the upper distance limit of 2200\,pc the
structures are low mass objects ($M=(1-6)\times10^{-4}\, 
\large({d\over100{\rm pc}}\large)^2$\mo)
which are not gravitationally virialized.
\ion{H}{I} 21cm line data towards the clouds show no prominent HI clouds. The
total \ion{H}{I} column densities for both structures are below
$N($HI$)\le2.1\times10^{20}$\,cm$^{-2}$, corresponding to $A_{\rm
V}\le0.2$\,mag, assuming a standard gas-to-dust ratio. IRAS 100\mue\ data towards
the structures show
also only low emission, consistent with low extinction. Unless there is unseen
cold dust associated with the structures this shielding is too low for the
structures to survive the interstellar radiation field for a long time. The
detection of 2 such structures in a rather limited sample of observations
suggests that they could be a rather common feature in the interstellar
medium, however, so far not recognized as such due to the weakness of their
lines and their small extent.
}
\maketitle

\section{Introduction}

H$_2$ is the most abundant molecule in the universe. Due to its missing
dipole moment and its high rotational constant it can only be observed 
directly in 
regions with high temperatures. The amount of molecular gas is however 
reasonably well estimated from observations of the second most abundant 
molecule, CO, if the metallicity of the gas is about solar (e.g. Dame et al.
\cite{dame:etal2001}). 

There has been much debate on how much molecular gas can be hidden from the
observer in form of small, cold molecular clouds (Pfenniger \& Combes
\cite{pfenniger:combes1994}; Gerhard \& Silk \cite{gerhard:silk1996}; Walker
\& Wardle \cite{walker:wardle1998}). Pfenniger \& Combes argue that 
 the molecular mass can be underestimated in the outer galaxy by as much as 
a factor 10,
if this gas has a fractal structure. Thus, in this form molecular gas could
account for all of the missing baryonic dark matter in our galaxy. These
so-called ``clumpuscules'' (Pfenniger \& Combes \cite{pfenniger:combes1994})
i) must be  small to escape detection, ii) must have high-column densities to
survive destruction and iii) must be dense to allow formation of H$_2$ within a
reasonable amount of time. Thus the question on baryonic dark matter in form
of small molecular clouds is directly linked to the question on the formation
and destruction of H$_2$ and CO.

In this paper the detection of a previously unobserved component
in the interstellar medium is described: 
molecular structures covering a very small area on
the sky located in an environment with very low hydrogen column densities.
These objects are ideal targets to study under which conditions CO and H$_2$
can survive in the interstellar medium.
The detection was made serendipitously during a search for molecular gas in
the tidal arms around the M\,81 group of interacting galaxies; the results of
this search will be described elsewhere.

\begin{figure*}
\rotatebox{-90}{\resizebox{9.5cm}{!}{\includegraphics{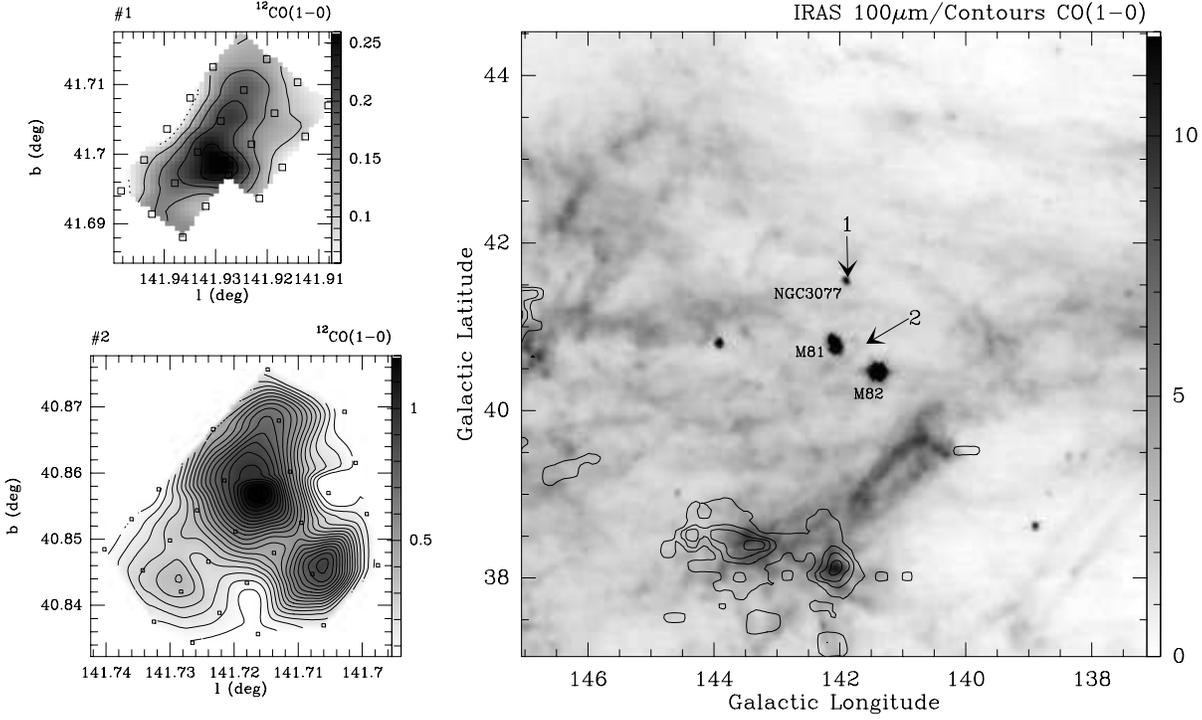}}}
\caption{The small-area molecular structures and their environment.
{\it On the right:}
Surrounding of the location were the two SAMS have been detected. The grey
scale is taken from the IRAS survey at 100\mue; the intensity scale is
displayed on the right of the box. Contours represent the CO (1$\to$0)
observations with the 1.2m CfA telescope observed by de Vries et al.
(\cite{devries:etal1987}); they are every 0.8\,K \kms\ starting at 0.8\,K \kms.
The point source near SAMS1 is  the galaxy NGC\,3077, the one near cloud
SAMS2 is M\,81 and the source below is M\,82. Arrows point to the centres of
the structures.
{\it On the left:}
Integrated CO (1$\to$0) intensity maps of the structures. The lowest contour
(dashed) is at 0.06\,K\,\kms\ ($3\sigma$), step size  between following
contours is 0.06\,K\,\kms. 
 Observed positions are marked by open squares.}
\label{overview}
\end{figure*}

\begin{table*}
\caption{Parameters for the molecular structures}
\begin{tabular}{l l l l l l l l l l}
\noalign{\hrule}
\noalign{\medskip}
 Number & $l$ & $b$ & transition & 
$T_{\rm A}^*$ & $rms$ & $v_{\rm lsr}$ & $\Delta v$ & $FWHM$ & $M$(H$_2$)$^{(2)}$\\
        &  (deg) & (deg)&         & 
(K)     &  (K)  & (\kms)    & (\kms)     & ($''$) & 
$10^{-4}\, \large({d\over100{\rm pc}}\large)^2$\mo\\
\noalign{\medskip}
\noalign{\hrule}
\noalign{\medskip}
\#1 & 141.9275 &     41.6970 &
CO (1$\to$0) & 0.142 & 0.008 & $+3.92\pm0.02$ & $0.75\pm0.04$ & 55 & 2\\
\#1 & & & 
CO (2$\to$1) & 0.034 & 0.009 & $+3.91\pm0.08$ & $0.87\pm0.15$ & $-$\\
\#2a & 141.7154 &    40.8557 & 
CO (1$\to$0) & 0.651 & 0.014 & $+5.38\pm0.01$ & $0.90\pm0.02$ & $60$ & 6\\
\#2a & & & 
CO (2$\to$1) & $0.24$ & 0.02 & $+5.35\pm0.02$ & $0.85\pm0.04$ & $-$\\
\#2b & 141.7078 & 40.8446 &
CO (1$\to$0) & 0.32 & 0.04 & $+5.00\pm0.03$ & $0.88\pm0.09$ & $25^{(1)}$ & 1\\
\#2b & & & 
CO (2$\to$1) & $0.37$ & 0.08 & $+4.88\pm0.04$ & $0.50\pm0.06$ & $-$\\
\noalign{\medskip}
\noalign{\hrule}
\noalign{\medskip}
\noalign{Remarks: Values for $T_{\rm A}^*$, $rms$, 
$v_{\rm lsr}$, and $\Delta v$ are derived from cloud averaged spectra.
1) Unresolved structure. 2) Mass is determined assuming
$X_{\rm CO}=0.6\times10^{20}$cm$^{-2}$(K km/s)$^{-1}$, 
$d$ is the distance to the cloud.}
\end{tabular}
\label{obstable}
\end{table*}

\section{Observations}

The observations were conducted in June 2001 and May 2002 with the IRAM 30\,m
telescope on Pico Veleta, Spain. In total, four  receivers were used, two tuned
to the $^{12}$CO (1$\to$0) line at 115\,GHz and two tuned to the $^{12}$CO
(2$\to$1) line at 230 GHz.
The observations were done with a wobbling secondary mirror with the
off-position separated by 200$''$ in azimuth from the on-position. The beam
size of the telescope is 22$''$ at 115\,GHz and 11$''$ at 230\,GHz. At the
beginning of the observing run the setup was optimized for extra-galactic
observations; i.e. for each of the 115\,GHz receivers a filterbank and an
autocorrelator spectrometer with a wide bandwidth and a low spectral
resolution were used. At 115\,GHz the filterbank had a velocity resolution of
2.6km/s and 512 channels, the autocorrelator had 0.8km/s resolution and 450
channels.

In total deep integrations on 25 individual positions were made in the tidal
arms of the M\,81 group of galaxies. During the observations I noticed at four
positions some spurious emission- and absorption-like features around $v_{\rm
lsr}=5$\,\kms\ which were only one channel wide with the setup chosen. To test
for a bad channel in the spectrometer or real spectral line the velocity
resolution was increased to 0.2\,\kms. Because the features turned out to
originate from two real interstellar clouds I started to map the two clouds
with a spacing of 20$''$ between two positions.

\section{Results \label{results}}

\begin{figure}
\rotatebox{-90}{\resizebox{4cm}{!}{\includegraphics{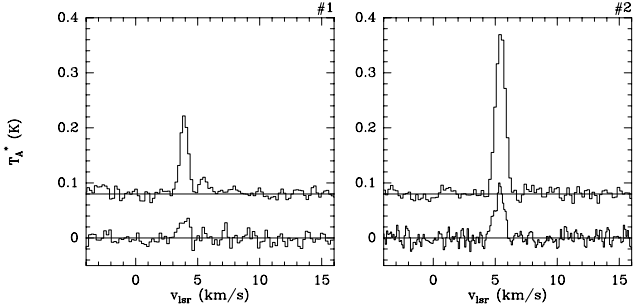}}}
\caption{CO (1$\to$0) and (2$\to$1) spectra towards the two clouds.
The  (1$\to$0) spectra have been shifted by 0.08\,K for better display.
The spectra are averaged over the whole cloud.}
\label{co_specs}
\end{figure}

The incomplete maps are displayed in Figure \ref{overview} together with an
IRAS 100\mue\ map of the surrounding of clouds. Cloud averaged spectra
reobserved with a higher spectral resolution are presented in
Fig.\,\ref{co_specs}, parameters are listed in Table \ref{obstable}. The
structures cover only very small areas on the sky. In analogy to the
tiny-scale atomic structures (TSAS) discussed in detail by Heiles
(\cite{heiles1997}) I will refer to them as small-area molecular structures
(SAMS) from here on.

SAMS1 appears as single elongated structure whereas SAMS2 can clearly be
separated into at least two structures, SAMS2a and SAMS2b. SAMS1, SAMS2a and
SAMS2b cover areas of $\ge9500$arcsec$^2$, 4800\,arcsec$^2$, and
1800\,arcsec$^2$, respectively. Due to the incomplete maps this is a lower
limit to the total area. Therefore I list the full widths at half maximum,
$FWHM$, of the SAMS, which are better determined. These are between 25$''$ and
60$''$. The average CO line temperatures of the SAMS are weak. Their
linewidths ($\Delta v\approx0.8$\,\kms) are lower than for galactic cirrus
clouds, for which the mean value is at 2.1\,\kms\ (Magnani et al.
\cite{magnani:etal1996}). If the clouds follow the same size-linewidth
relation found for many types of molecular clouds, $\big({\Delta
v\over\kms}\big)^2\approx {FWHM\over \rm pc}$ (e.g Larson \cite{larson1981};
Heithausen \cite{heithausen1996}), the clouds must be closer than 2200\,pc,
i.e. they are galactic objects. 

Assuming  $X_{\rm CO}=0.6\times10^{20}$\,cm$^{-2}$(K\,\kms)$^{-1}$ (de Vries
et al. \cite{devries:etal1987}) the average H$_2$ column densities and 
the masses of 
the SAMS correspond to $N$(H$_2)=(0.7-3.7)\times10^{19}$\,cm$^{-2}$ and
$M($H$_2)=(1-6)\times10^{-4}({d\over100{\rm pc}})^2$\,\mo, where $d$ is the distance to
the structures. With such low masses the structures are not gravitationally
virialized; even at the upper distance limit the virial mass, $M_{\rm
vir}=210\,{\rm M_\odot}*{\Delta v^2\over\kms}*{FWHM\over {\rm pc}}\approx4
({d\over100{\rm pc}})$\,\mo, is orders of magnitude too large. Note, however,
that these parameters are only valid if galactic conversion factors are
applicable, which has to be tested by further observations.

The SAMS are also detected in the CO (2$\to$1) line at various positions.
Figure \ref{co_specs} shows the cloud averaged spectra. The corresponding
values are listed in Table \ref{obstable}. Due to the undersampling at
230\,GHz and the inhomogeneous rms caused by variable summer conditions during
the observations, the (2$\to$1) data provide only limited structural
information on the clouds. 

The two newly detected structures are located in a low column density 
environment. Fig.\,\ref{overview} gives an overview over the region. The
positions are more than two degrees away from the nearest molecular
high-latitude clouds studied by de Vries et al. (\cite{devries:etal1987}). The
level of 100\mue\ emission observed by IRAS is 2\,MJy\,sr$^{-1}$. Using the
dust emissivity found for galactic cirrus clouds $I(100\mu{\rm m})/N({\rm
H})=1.0\times10^{-20}$\,MJy\,sr$^{-1}$\,cm$^2$ (Heithausen \& Mebold
\cite{heithausen:mebold1989}) a total column density of the local
gas of $N$(H)=$2\times10^{20}$cm$^{-2}$ is derived on a 4$'$ scale.

\begin{figure}
\rotatebox{-90}{\resizebox{6cm}{!}{\includegraphics{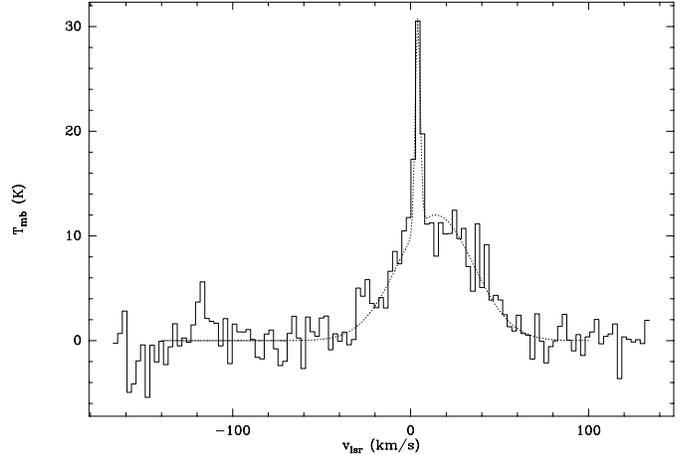}}}
\caption{Combined Effelsberg and VLA spectrum of the \ion{H}{I} 21cm line
towards SAMS1. The dotted line is a fit to the data.}
\label{hi_specs}
\end{figure}

To derive the column density of atomic gas in the direction of SAMS1, published
\ion{H}{I} 21cm line  observations obtained with the VLA (Walter \& Heithausen
\cite{walter:heithausen1999}; Walter et al. \cite{walter:etal2002}) were
corrected for missing zero-spacings using data obtained with the Effelsberg
100\,m telescope. The resulting angular resolution was $19''\times16''$, the
velocity resolution 2.6\,\kms. No evidence for an \ion{H}{I} cloud associated
with SAMS1 is apparent in the combined VLA/Effelsberg data. The spectrum
towards the maximum of SAMS1 shows two velocity components: one broad line
coming from the extragalactic tidal arms of the M\,81 system and one narrow
component coming from local galactic material (s. Fig. \ref{hi_specs}). The
local gas component at $v_{\rm lsr}=4.0$\,\kms\ has a \ion{H}{I} column
density of only $N$(HI)=$(1.3\pm0.1)\times10^{20}$cm$^{-2}$.

For SAMS2  only information on the local atomic hydrogen on a $9'$ scale exist
from observations with the Effelsberg 100\,m telescope (de Vries et al.
\cite{devries:etal1987}). The column density is found to be
$N$(HI)=2.1$\times10^{20}$\,cm$^{-2}$, consistent with the value derived above
from the IRAS data. 

In summary, the column densities along the line of sights towards the
SAMS are very low,
corresponding to only low visual extinction of $A_{\rm V}\le0.2$\,mag, assuming
a standard gas-to-dust ratio. One should however keep in mind that
with the information presently available, the existence of very cold dust
associated with the SAMS cannot be excluded.

\section{Discussion}

Are we faced with an unknown feature of the interstellar
medium or are we just observing the very low mass end of ordinary cirrus
clouds? In this section I argue that the structures are different from galactic
molecular cirrus clouds in several ways.

The clouds described in this paper are so small and weak that they are missed
by any galactic survey conducted so far (e.g. Heithausen et al.
\cite{heithausen:etal1993}; Hartmann et al. \cite{hartmann:etal1998}). E.g. in
the beam of the CfA 1.2m telescope, the telescope which did  most of the
inventory of molecular clouds in 
the Milky Way (Dame et al. \cite{dame:etal2001}), SAMS1 would
appear as a 1.6\,mK feature, orders of magnitude below the sensitivity of any
galactic survey conducted with this telescope.

Besides their compactness SAMS differ from molecular cirrus clouds  because
they reside in a low column density environment which provides little
shielding against the interstellar radiation field. For cirrus clouds the good
correlation between CO, \ion{H}{I} and IRAS 100\mue\ data has been used to
determine the total hydrogen column density and calibrate the $X_{\rm CO}$
factor (e.g. de Vries et al. \cite{devries:etal1987}). In fact, all molecular
cirrus clouds from a large-scale CO survey at high galactic latitudes were
seen in the IRAS 100\mue\ maps, although no one-to-one correlation between CO
and the 100\mue\ intensity exists (Heithausen et al.
\cite{heithausen:etal1993}). For the SAMS described here there is no
counterpart in either the IRAS data (though the angular resolution of IRAS
might be possibly too coarse to detect them) or in high angular resolution
\ion{H}{I} data. 

Unless SAMS are asocciated with very cold dust which is unseen by IRAS,
the extinction of the structures is too low to provide efficient
shielding for CO to survive. Modelling of the photodissociation of CO requires
visual extinction of $\ge1$\,mag or total hydrogen column densities of 
$\ge10\times10^{20}$\,cm$^{-2}$ for the existence of CO (e.g. Viala et al.
\cite{viala:etal1988}; van Dishoeck \& Black \cite{vandishoeck:black1988}; 
van Dishoeck \& Blake \cite{vandishoeck:blake1998}). Below that limit CO might
exist but with a very limited lifetime.

On the other hand, the formation time for H$_2$ in a diffuse atomic
environment is rather long. Pironelli et al. (\cite{pironelli:etal2000})
calculate that for a shielding of $A_{\rm V}=1$\,mag and
$n$(H$_2$)=100\,cm$^{-3}$ it takes almost 10$^8$ years for H$_2$ to built up
from atomic hydrogen in large amounts. This time is much longer than the
dynamical timescale of the  observed structures which is $FWHM/\Delta
v\approx2\times 10^4\times ({d\over100{\rm pc}})$\,years.

\section{Conclusions}

In this Letter I have described the detection of a previously unobserved
component of the interstellar medium: small-area molecular structures embedded
in a very low column density environment. Aiming at the detection of molecular
clouds in extragalactic tidal arms, the observations were not set up to detect
local clouds, thus they are unbiased. Based on their location 
in a low column density environment
with insufficient shielding against the
interstellar radiation field, the clouds should not exist for very long.
Whether sufficient shielding is provided by 
very cold dust remains to be tested by bolometer observations. 

Whether these two structures were picked up just by chance before they are
dissolved or whether their detection indicates that H$_2$ and CO can survive
with less shielding than previously adopted is an open question.
Because two structures were discovered in a rather 
limited sample of observations, one could  speculate that such
structures might be frequent in the interstellar medium. This explanation is
also supported by the detection of H$_2$ in many lines of sight through
diffuse galactic clouds in the spectra of  distant quasars (e.g. Richter et
al. \cite{richter:etal2001}). Due to their weakness and compactness these
structures will however be missed by most observations.

Whether or not  SAMS are related to the tiny-scale atomic structures (TSAS,
Heiles \cite{heiles1997}) is unclear. TSAS have been found in \ion{H}{I} 21cm
line observations. They are thought to be predominantly made of atomic gas and
 have sizes of about 25\,AU (Diamond et al. \cite{diamond:etal1989}).  SAMS
described here are molecular and have unknown sizes, due
to the unknown distance. If we adopt for the moment that they have the same
sizes as the TSAS, they would be as close as 0.5\,pc. 
At that distance they would only live for a few hundred years.
Whether SAMS could be
related to the sodium absorption line clouds which were found to be
non-coincident with \ion{H}{I} clouds (Lilienthal \& Wennmacher
\cite{lilienthal:wennmacher1990}) remains to be tested by sensitive CO
observations. If so, these observations will provide the crucial 
distance information for SAMS.

To estimate the frequency of SAMS, it will be worthwhile to check existing CO
observations made towards extragalactic objects for the presence of similar
absorption- or emission-like features. 
Clearly, more information on SAMS are needed before one could draw more firm 
conclusions on their nature. 

\begin{acknowledgement}
I thank J\"urgen Ott for providing Effelsberg HI data, Fabian Walter for
providing the VLA data,  and Axel Wei\ss\ for doing the zero-spacing
correction. I thank Ulrich Mebold, Fabian Walter and Peter Kalberla for 
critically reading the manuscript.
\end{acknowledgement}

{}

\end{document}